\documentclass[12pt,preprint]{aastex}

\begin{document}

\newcommand{\vdag}{(v)^\dagger}
\newcommand{\opeqn}{\begin{equation}}
\newcommand{\cleqn}{\end{equation}}

\title{ How Cumbersome is a Tenth Order Polynomial?: \\ 
 The Case of Gravitational Triple Lens Equation} 

\author{Sun Hong Rhie (UND)}

\begin{abstract}
Three point mass gravitational lens equation is a two-dimensional vector 
equation that can be embedded in a tenth order analytic polynomial equation 
of one complex variable, and we can solve the one variable equation on the 
source trajectories using recipies for Fortran or $C$ (portable for $C$++ or 
$C_{jj}$) in Numerical Recipes, or using packages such as Mathemetica, 
Matlab, etc. This ready solvability
renders fitting microlensing light curves including triple lenses a normal 
process, and such was done in a circumbinary planet fit for MACHO-97-BLG-41.  
Subsequently, there was a claim that converting the triple lens equation into 
the analytic equation was rather cumbersome, and the impressionable judgement 
has caused an effect of mysterious impedance around the perfectly tractable 
lens equation. There are judgements.  Then, there is nature. 
We looked up for one of the quantities of highest precision measurements: 
electron $g$-factor correction $a_e \equiv g/2-1$. The current best experimental 
values of $a_e$ agree to eight significant digits with the theoretical value, 
and the theoretical calculation involves more than one thousand Feynman diagrams 
-- many orders of magnitude messier than the triple lens equation coefficients.  
We seem to have only choice to be compliant to nature and its appetite for 
elegant mess and precision numerics.  In fact, the triple lens equation 
coefficients take up less than a page to write out and are presented here 
for users' convenience.

\end{abstract}


\clearpage
\newpage

\section {Gravitational Triple Lens Equation} 

The lens equation of a set of three gravitationally bound point masses 
is written with three (real) relative mass parameters, $\epsilon_1$, 
$\epsilon_2$, and $\epsilon_3$, and three  complex parameters for their 
2-dimensional positions. If $z$ is the position of an image and $\omega$
is the position of its source,  the lens equation reads as follows.
\opeqn
   \omega = z - {\epsilon_1\over \bar z - \bar x_1}
              - {\epsilon_2\over \bar z - \bar x_2}
              - {\epsilon_3\over \bar z - \bar x_3}
       \ \equiv \ z - f(\bar z; \bar x_j)
\label{eqTrilens}
\cleqn

The mass parameters are subject to a constraint
$\epsilon_1 + \epsilon_2 + \epsilon_3 =1$ where $1 = M$ is the total
mass, and the position variables include coordinate degrees of freedom: 
\ two degrees of freedom for translation and one degree of freedom for 
roation in the two-dimensional lens plane. 
We choose a coordinate system  such that the position of a mass,
$x_1$, and the center of mass of the other two elements, $x_4$,
define the lens axis along the real axis of the complex plane.
\opeqn
  (\epsilon_2 + \epsilon_3) ~x_4 = \epsilon_2 x_2 + \epsilon_3 x_3
\cleqn
Then, $x_1$ and $x_4$ are real, and the triple lens system is completely
specified by a set of five parameters.
\opeqn
 \ell = |x_1 - x_4| \ ; \  \
 \ell^{\prime} = |x_2 - x_3| \ ;  \  \
 \gamma = [0, \pi] \ ; \  \
 \epsilon_2 = (0, 1) \ ;  \ \
 \epsilon_3 = (0, 1) \
\cleqn
The angle $\gamma$ is the angle between $(x_2 - x_3)$ and the lens axis.
In order to find the images of a given source, we need to solve the lens
equation. 

When the lensing system and a source star are far apart, there 
are four images of the source star, which we can verify easily for a source
at  $\omega = \infty$.   From the lens equation (\ref{eqTrilens}), we
find that the images of the source at $\omega = \infty$ are
at the position of the source, $z = \infty$, and at the three lens
positions, $z = x_1, x_2$, and $x_3$.  The magnifications of the images
at the lens positions are $0$, and they are hardly images in any practical
sense because there are no photon fluxes related to the image positions.
In other words, there is only one physical image for a source at $\infty$
-- namely, the image of the unmagnified source star.
However, if we move the source toward the lens system, the images at the
lens positions move away from the lens positions, and they do have
non-vanishing photon fluxes.  Thus, we freely speak of image positions with
zero magnification as the continuity limit so that we don't have to cut out
the lens positions from the image space. In practice, $\infty$ is a  large 
distance limit, which may be no more than a few hundred au in any physical 
relevance of Galactic lensing.

If we consider a source trajectory away from the caustic regions, the images
form four smooth curves anchored at the {\it base points} of 
$z = x_1, x_2, x_3$, and $\infty$ which are the image positions of the
source at $\omega = \infty$.  The complex lens plane can be considered a large
two sphere with one point at $\infty$, and then, the four smooth disjoint 
image curves are loops with fixed points at the {\it base points}. 
The image curves can be calculated esaily from 
the {\it base points} using Newtonian transportation method (the algorithm 
can be found in Numerical Recipes). However, the physical interest of lensing
lies in caustic regions where the lensing signals are most obvious, and this
makes the Newtonian method useless in microlensing business. The large lensing 
signals arise where the inverse Jacobian determinant of the lens equation is
large, and the large Jacobian determinant makes Newtonian transportation method 
unstable and unusable. The method of infinitesimal corrections based on 
derivative values is valid where the Jacobian determinant is practically 
finite and the mapping is non-degenerate. Laguerre's method (see Numerical 
Recipes) is based on complex analysis and uses first and second derivatives 
of the polynomial to help converge to the solutions from initial values 
which one is free to guess.    
  
Another complication of the caustic region is that the size 
(which may be $1-10\mu$as)
of the source star can not be ignored when the source star crosses a caustic 
curve. In fact, the luminosity profile of the source star can manifestly 
affect the shape of the light curve, and the luminosity profile dependence of
the light curve shape during a line caustic crossing lasts $\sim 3$ stellar 
radius crossing times \citep{rb99}.  In order to incorporate the finite size 
effects, one can {\it locally} pixelize the image plane ($z$-plane) in the 
neighborhood of the critical curve near the images and count up 
the image pixels that are mapped into the source disk \citep{br96}. 
The luminosity profile of the source can be easily incorporated by weighting
each image pixel by the luminosity shape function value at the corresponding 
source position.  The pixel size can be adjusted for a desired resolution. 
In the ray shooting method which has been designed mainly to
handle a system of a large number of lensing elements solving whose lens 
equation is impractical if not impossible,  the entire lens plane
is pre-pixelized and each pixel in the image plane is tested whether the center 
of the pixel is mapped into the area defined by the source disk in the 
source plane.  The ray shooting method is inefficient for fitting light 
curves of low-multiplicity point mass lenses as microlensing planet systems.   

The astro-ph version of \citet{gns98} conveys an impression that solving the 
lens equation by finding the roots of the 10-th order polynomial analytic 
equation is slower than using the ray shooting method for triple lens light 
curve fitting. However, Gaudi (private communication) recently informed us 
that the comment on the calculation speed was a comparison between binary 
lens and triple lens but not a comparison between root finding 
method and ray shooting method for the triple lens equation.

\section{The Tenth Order Polynomial Equation}

The lens equation is an explicit function from an image position $z$ to its
source position $\omega$, and $\omega (z, \bar z)$ is a genuine real function, 
namely a function of both $z$ and $\bar z$.  However, the $z$-depdence is linear, 
and this simplicity is behind the analyticity of the differential behavior which
can be completely described by one analytic function 
$\kappa \equiv \partial_z \bar\omega$.
The linearity in $z$ also makes it easy to find an analytic equation where the
lens equation is embedded.    Using $z - \omega = f(\bar z; \bar x_j)$
and $\bar z - \bar\omega  = f(z; x_j)$,
\opeqn
 z - \omega = f\left(f(z;x_j)+\bar\omega;\bar x_j \right)
\label{eqAAnal}
\cleqn
If we let $H \equiv z_1 z_2 z_3$ and
$G\equiv \epsilon_1 z_2 z_3 + \epsilon_2 z_3 z_1 + \epsilon_3 z_1 z_2
 = \sum_{\rm cyc} \epsilon_i z_j z_k$,
then $f = G/H$, and it is simple to see that equation (\ref{eqAAnal}) is
a tenth order polynomial equation.
If we let $\bar\omega_j \equiv \bar\omega - \bar x_j$,
\opeqn
 0 = (z-\omega)(G+\bar\omega_1 H)(G+\bar\omega_2 H)(G+\bar\omega_3 H)
    -  \sum_{\rm cyc} H \epsilon_i (G+\bar\omega_j H)(G+\bar\omega_k H)
\label{eqApolynomial}
\cleqn
An analytic polynomial equation has the same number of solutions as the order 
(see any textbook on complex variable or mathematical physics), there can be 
up to ten images in a triple lensing.   
There are only four images for $\omega = \infty$, the number of images 
changes by two at a caustic crossing, and the caustic curves form 
heierachical structures of domains for high multiplicity images.  
There are triple lenses with domain ${\cal D}^3$ the sources therein produce
ten images, the maximum possible number of images \citep{rh97}. As we 
repeatedly emphasized in \citet{rh97}, the triple lens equation is equivalent 
to the tenth order polynomial analytic equation only in the domain 
${\cal D}^3$, and the statement on the equivalence in section 3 of
\citet{gns98} seems to be a misquote which can mislead the readers to 
think that the two equations are equivalent everywhere.

In order to fit a triple microlensing light curve, we need to solve the tenth
order analytic equation (and select the image solutions that satisfy the lens
equation).  This can be done numerically using root finders
available in the literature, and we only need to type in the coefficients.  
The coefficients may appear to be cumbersome as declared in \citet{gns98},  
and it is indeed the case if we, for example,
calculate the coeffients using Mathematica.  The output of an algebraic 
computing package is (unnecessarily) messy even for the binary lens equation.  
Thus, it is useful to group (or not to unfold) the coefficients, which is 
a natural intermediate process in hand calculations. 

In the center of mass system, $\epsilon_j x_j =0$, $H$ and $G$ are written
with four coefficient functions, $a, b, c$, and $d$: \
$ H =  z^3 + a z^2 + b z + c$ \ and $G =  z^2 + a z + d $,  where
$a \equiv -(x_1+x_2+x_3),  \  b \equiv x_1 x_2 + x_1 x_3 + x_2 x_3,  \
 c \equiv - x_1 x_2 x_3$,  and
$ d \equiv \sum_{\rm cyc} \epsilon_i x_j x_k$.
If we let $a_\omega \equiv \bar\omega_1 + \bar\omega_2 + \bar\omega_3$, \
$b_\omega \equiv \bar\omega_1 \bar\omega_2 + \bar\omega_2 \bar\omega_3
+ \bar\omega_3 \bar\omega_1$, \
$c_\omega \equiv \bar\omega_1 \bar\omega_2 \bar\omega_3$, and
$d_\omega \equiv \sum_{\rm cyc} \epsilon_i \bar\omega_j \bar\omega_k$,
then equation (\ref{eqApolynomial}) becomes
\opeqn
 0 = G^3 (z - \omega)
    + G^2 H ((z - \omega) a_\omega + 1)
    + G H^2 ((z - \omega) b_\omega + a_\omega - \bar\omega)
    + H^3 ((z - \omega) c_\omega + d_\omega)
\cleqn
If we let $ G^3 \equiv H_{0k} z^k ~(k\le 6) $, \
            $ G^2 H \equiv  H_{1k} z^k ~(k\le 7) $,  \
             $ G H^2 \equiv   H_{2k} z^k ~(k\le 8) $, \
         and $ H^3  \equiv H_{3k} z^k ~(k\le 9)  $, \  the equation becomes
\opeqn
 0 \ = \ \sum_{k=1}^{10} \  {\rm cff}~(k) \  z^k
\cleqn
where the polynimial coefficients are
\begin{itemize}
\item
$
 {\rm cff}~(k) \ =  \left(\ H_{0 k-1} + H_{1 k-1}~ a_\omega + H_{2 k-1}~ b_\omega
                  + H_{3 k-1}~ c_\omega \right)  \\  
      - \left(H_{0k}~\omega + H_{1k}~ (\omega a_\omega -1) 
            + H_{2k}~ (\omega b_\omega + a_\omega - \bar\omega)
                  + H_{3k}~ (\omega c_\omega + b_\omega) \right)
$
\end{itemize}
The coefficients $H_{ij}$ are polynomials of $a, b, c$, and $d$ where the 
polynomial coefficients are simple combinatoric integers.

\begin{itemize}
\item
$ H_{39} = 1;  \
  H_{38} = 3 a; \
  H_{37} = 3 b + 3 a^2;  \
  H_{36} = 3 c + 6 a b + a^3;   \
  H_{35} = 6 a c + 3 b^2 + 3 a^2 b;  \
  H_{34} = 6 b c + 3 a^2 c + 3 a b^2; \
  H_{33} = 3 c^2 + 6 a b c + b^3; \
  H_{32} = 3 a c^2 + 3 b^2 c;  \
  H_{31} = 3 b c^2;  \
  H_{30}= c^3$.

\item
$      H_{28} = 1; \
       H_{27} = 3 a ; \
       H_{26} = d + 2 b + 3 a^2  ; \
       H_{25} = 2 a d + 4 a b + a^3 + 2 c  ; \
       H_{24} = 2 d b + d a^2 + 4 a c + 2 a^2 b + b^2   ; \
       H_{23} = 2 d c + 2 d a b + 2 a^2 c + a b^2 + 2 b c  ; \
       H_{22} = 2 c a d + d b^2 + 2 a b c + c^2   ; \
       H_{21} = 2 b c d + a c^2   ; \
       H_{20}= c^2 d $

\item
$      H_{17} = 1                  ; \
       H_{16} = 3 a                  ; \
       H_{15} = 2 d + 3 a^2 + b         ; \
       H_{14} = 4 a d + a^3 + 2 a b + c   ; \
       H_{13} = d^2 + 2 a^2 d + 2 b d + b a^2 + 2 a c  ; \
       H_{12} = a d^2 + 2 a b d + 2 c d + c a^2  ; \
       H_{11} = b d^2 + 2 a c d   ; \
       H_{10} = c d^2  $

\item
$      H_{06} = 1     ; \
       H_{05} = 3 a       ; \
       H_{04} = 3 d + 3 a^2      ; \
       H_{03} = 6 a d +  a^3     ; \
       H_{02} = 3 d^2 + 3 a^2 d^2   ; \
       H_{01} = 3 a d^2   ; \
       H_{00} = d^3   $
\end{itemize}

\subsection{Comments}

An interested party may download the source file of this manuscript 
to avoid tying the coefficients. We encourage to check the coefficients,
however. It is a quick exercise once one adopts the poor person's calculation
with pencil and paper as shown above; also with the free biocomputer
which is harder to hack either internally or externally barring the long
term process of brainwashing.  
We also found it useful to test the symmetric cases in rh97 whose 
image solutions behavior (for example, the number of images) is known.   
The critical curve is obtained by solving $\kappa = e^{i2\varphi}$ which
is an eighth order polynomial equation.  The caustic curve is obtained by 
applying the lens equation to the critical curve solution.   It is fine to 
use $\varphi = [0, 2\pi)$ as the parameter for equal interval sampling.
Let $\delta = 2\pi/N$ for $N$ not too large and observe the intervals
(or speeds) of the solutions on the critical curve and the caustic curve.
Note especially the density of the solutions around the cusps of the 
caustic curve. It is worth pausing for a moment counting the relative numbers 
of the solutions on the stellar caustic and planetary caustics for planet 
systems lenses.  

For the measurements and theoretical calculations of the anomalous magnetic 
moment of the electron, we have consulted Peskin and Schroeder (1999) 
and Kinoshita (1990). 
We have considered drawing all the Feynman diagrams
to lay out the degree of lengthy squiggly messiness but given up, and 
our misadventure may be considered an indirect testimony of the degree 
of mud wrestling the exquisite anomalous magnetic moment requires.       

Barring the notion that lensing community may have been chosen to be dealt 
with laxed scrutiny of nature, we have no doubt that we have no luxury to 
complain about a bit of algebra we encounter in lensing. In fact, we 
find it a cherished treasure that low multiplicity point lenses are exactly 
solvable and their light curves can be reconstructed and interpreted
without ambiguities.  Microlensing events do share the transiency with 
the scattering events in accelerator particle physics. The both need high 
resolution data for minute rare prized signals and complete data to 
interpret them. The both rely on methodical analyses building from the     
simpler and dominant events to more rare events. Thus, it is
important to have  a homogeneous and comprehensive data set of 
microlensing events where consistencies can be tested within
in order to find microlensing planets.  The best bet for such data set
is microlensing from space. Accelerator particle physics of the last
century is at the foundation of the Standard Model or the Theory of 
Matter.  We expect that space microlensing planet search will lay
a foundation of extrasolar planet physics within a few years of operation
of a small space telescope \citep{gest}. In the comprehensive data set,
the so-called high magnification events with only stellar caustic signals
will form an independent data set that can be used for a consistency 
check of the interpretation of the planetary light curves as a subset.

This note is based on the work with D. Bennett for Bennett et al. (1999).


\clearpage

\end{document}